\documentclass[conference]{IEEEtran}
\IEEEoverridecommandlockouts

\usepackage{float}
\usepackage{balance}
\usepackage{cite}
\usepackage{amsmath,amssymb,amsfonts}
\usepackage{algorithmic}
\usepackage{graphicx}
\usepackage{soul}
\usepackage{textcomp}
\usepackage[dvipsnames]{xcolor}
\usepackage[inline]{enumitem}
\usepackage{subcaption}
\usepackage{calc}
\usepackage{float}
\usepackage{calc}
\usepackage{hyperref}
\usepackage[disable]{todonotes}
\usepackage{courier}

\usepackage{listings}
\usepackage{xcolor}

\lstdefinestyle{mystyle}{
    commentstyle=\color{gray}\ttfamily,
    numberstyle=\tiny\color{darkgray},
    stringstyle=\color{purple},
    basicstyle=\footnotesize,
    breakatwhitespace=false,         
    breaklines=true,                 
    captionpos=b,                    
    keepspaces=true,                 
    numbers=left,                    
    numbersep=5pt,                  
    showspaces=false,                
    showstringspaces=false,
    showtabs=false,                  
    tabsize=1,
    upquote=true,
    frame=single,
    morekeywords={as, include, printf, a}
}
\makeatletter
\def\minted@breakchar{}
\makeatother

\newcommand{\ourvspace}{\vspace{-0.4em}}

\newcommand{\class}[1]{\textcolor{RoyalBlue}{\texttt{#1}}}

\newcommand{\inst}[1]{{\fontfamily{pcr}\selectfont\small #1}}

\def\BibTeX{{\rm B\kern-.05em{\sc i\kern-.025em b}\kern-.08em
    T\kern-.1667em\lower.7ex\hbox{E}\kern-.125emX}}

\begin{document}

\bstctlcite{IEEEexample:BSTcontrol}

\title{
PROV-AGENT: Unified Provenance for Tracking \\AI Agent Interactions in Agentic Workflows
\thanks{
Cite this paper as: 
R. Souza, A. Gueroudji, S. DeWitt, D. Rosendo, T. Ghosal, R. Ross, P. Balaprakash, R. F. da Silva, ``PROV-AGENT: Unified Provenance for Tracking
AI Agent Interactions in Agentic Workflows,'' 
Proceedings of the 21st IEEE International Conference on e-Science (e-Science), Chicago, IL, USA, 2025.\\
{\tiny
This manuscript has been authored by UT-Battelle, LLC, under 
    contract DE-AC05-00OR22725 with the US Department of Energy (DOE). 
    The publisher, by accepting the article for publication, acknowledges 
    that the U.S. Government retains a non-exclusive, paid up, irrevocable, 
    worldwide license to publish or reproduce the published form of the 
    manuscript, or allow others to do so, for U.S. Government purposes. 
    The DOE will provide public access to these results in accordance 
    with the DOE Public Access Plan 
    (http://energy.gov/downloads/doe-public-access-plan).
}
}
\ourvspace{}
}

\author{
\IEEEauthorblockN{
    Renan Souza\IEEEauthorrefmark{1},
    Amal Gueroudji\IEEEauthorrefmark{2},
    Stephen DeWitt\IEEEauthorrefmark{3},
    Daniel Rosendo\IEEEauthorrefmark{1},
    Tirthankar Ghosal\IEEEauthorrefmark{1},\\
    Robert Ross\IEEEauthorrefmark{2},
    Prasanna Balaprakash\IEEEauthorrefmark{4},
    Rafael Ferreira da Silva\IEEEauthorrefmark{1}
}
\IEEEauthorblockA{
    \IEEEauthorrefmark{1}National Center for Computational Sciences, Oak Ridge National Lab, Oak Ridge, TN, USA\\
    \IEEEauthorrefmark{2}Mathematics and Computer Science Division, Argonne National Laboratory, Lemont, IL, USA\\
    \IEEEauthorrefmark{3}Computational Sciences and Engineering Division, Oak Ridge National Lab, Oak Ridge, TN, USA\\
    \IEEEauthorrefmark{4}Computer Science and Mathematics Division, Oak Ridge National Lab, Oak Ridge, TN, USA\\
}
    
}

\maketitle


\begin{abstract}
Large Language Models (LLMs) and other foundation models are increasingly used as the core of AI agents. 
In \textit{agentic workflows}, these agents plan tasks, interact with humans and peers, and influence scientific outcomes across federated and heterogeneous environments.
However, agents can hallucinate or reason incorrectly, propagating errors when one agent’s output becomes another’s input. 
Thus, assuring that agents’ actions are transparent, traceable, reproducible, and reliable is critical to assess hallucination risks and mitigate their workflow impacts.
While provenance techniques have long supported these principles, existing methods fail to capture and relate agent-centric metadata such as prompts, responses, and decisions with the broader workflow context and downstream outcomes.
In this paper, we introduce PROV-AGENT, a provenance model that extends W3C PROV and leverages the Model Context Protocol (MCP) and data observability to integrate agent interactions into end-to-end workflow provenance. Our contributions include: (1)~a provenance model tailored for agentic workflows, (2)~a near real-time, open-source system for capturing agentic provenance, and (3)~a cross-facility evaluation spanning edge, cloud, and HPC environments, demonstrating support for critical provenance queries and agent reliability analysis.

\end{abstract}

\section{Introduction}

The integration of foundation models, often referred to as Large ``X" Models (LxMs), into computational workflows is rapidly advancing across scientific and industrial domains~\cite{fettke2025llm}. These models excel in language, vision, time-series, and robotics tasks, driving innovation in genomics, chemistry, and manufacturing. This shift has driven the emergence of \textit{agentic workflows}, where autonomous agents make decisions, plan tasks, and coordinate with humans and other agents. These agents operate in dynamic environments across heterogeneous computing platforms, including edge devices, cloud systems, and high-performance computing (HPC). Unlike traditional workflows with static, deterministic paths~\cite{suter2025terminology}, agentic workflows are non-deterministic, shaped by near real-time data, adaptive decisions, and evolving interactions~\cite{pauloski2025empowering, da2025grassroots}. They often display dynamic, cyclic behavior, where agent outputs inform subsequent decisions or feedback loops.

Although Artificial Intelligence (AI) agents offer capabilities for automating complex processes, they introduce new challenges for transparency, reproducibility, and reliability. They may generate hallucinated or incorrect outputs, especially when relying on generative models, which can propagate through the workflow, compounding errors and making it difficult to assess the correctness of the results~\cite{agent_smith}. The risks are amplified in workflows where agent decisions influence other agents or downstream tasks, potentially affecting scientific conclusions or operational outcomes.
%
Provenance data management has long played a central role in providing for
such transparency, reproducibility, and reliability in computational workflows~\cite{10678731}. 
However, traditional provenance approaches are not designed to capture the intrinsic dynamics of modern AI agents. Provenance data must not only capture the data flow and task execution history but also represent the reasoning processes, model invocations, and contextual information that drive agent decisions. This level of detail enables rigorous root cause analysis when unexpected or erroneous behavior occurs. For example, understanding how a surprising result was produced requires tracing back through multiple agent interactions, prompts, responses, and intermediate computations.

A unified provenance graph that considers AI agent actions as first-class components, on par with traditional workflow tasks, enables comprehensive traceability and analysis. This structure supports critical queries such as: (1)~\textit{What specific input data led an agent to make a particular decision?} (2)~\textit{How did an agent’s decision influence the control or data flow within the workflow?} (3)~\textit{Which downstream outputs were affected by a specific agent interaction?} (4)~\textit{Where did erroneous data originate, and through which agents decisions or workflow tasks did it propagate?} These questions are essential for interpreting results, debugging workflows, and improving agent performance through better prompts and model tuning.

In this paper, we build on foundational efforts in workflow provenance research to introduce a framework that captures both traditional workflow metadata and AI agent interactions. Our contributions are threefold: (1)~\textbf{PROV-AGENT}, a provenance model that extends the W3C PROV~\cite{W3CPROV} standard and incorporates concepts from the Model Context Protocol (MCP)~\cite{mcp} to represent agent actions and their connections to data and workflow tasks; (2)~an open-source system~\cite{flowcept} for data observability and runtime agentic provenance capture in workflows; and (3)~a preliminary evaluation with a cross-facility agentic workflow involving edge devices, cloud services, and HPC systems. 

\section{Background and Related Work}
\label{sec:background}

\subsection{Provenance for Tracking AI Agents in Dynamic Cross-Facility Workflows}
\label{sec:agents_mcp}

Agentic workflows are emerging as a new paradigm in scientific computing, where autonomous AI agents are integrated into complex, multi-step processes. These agents, often powered by foundation models such as LLMs, take on responsibilities traditionally handled by humans or static scripts. They interpret data, make decisions, and adaptively steer workflow execution. To support the development and orchestration of such agentic workflows, a variety of frameworks have emerged. 
For instance, 
LangChain~\cite{topsakal2023creating, auffarth2023generative}, AutoGen~\cite{wu2023autogen}, LangGraph~\cite{wang2024agent}, Academy~\cite{pauloski2025empowering}, and CrewAI~\cite{crewai} support multi-agent systems that interact through prompt exchanges, calls to foundation models typically hosted by AI service providers in the cloud (e.g., OpenAI, SambaNova), and shared context. These frameworks support MCP~\cite{mcp}, which is emerging as a standard in academia and industry. MCP defines core agentic AI development concepts, including tools, prompts, resources, context management, and agent–client architecture that can communicate with external sources, such as knowledge bases or web pages, for Retrieval-Augmented Generation (RAG)~\cite{gao2023retrieval} to dynamically augment prompts.

\begin{figure}[!ht]
  \centering
  \includegraphics[width=\columnwidth]{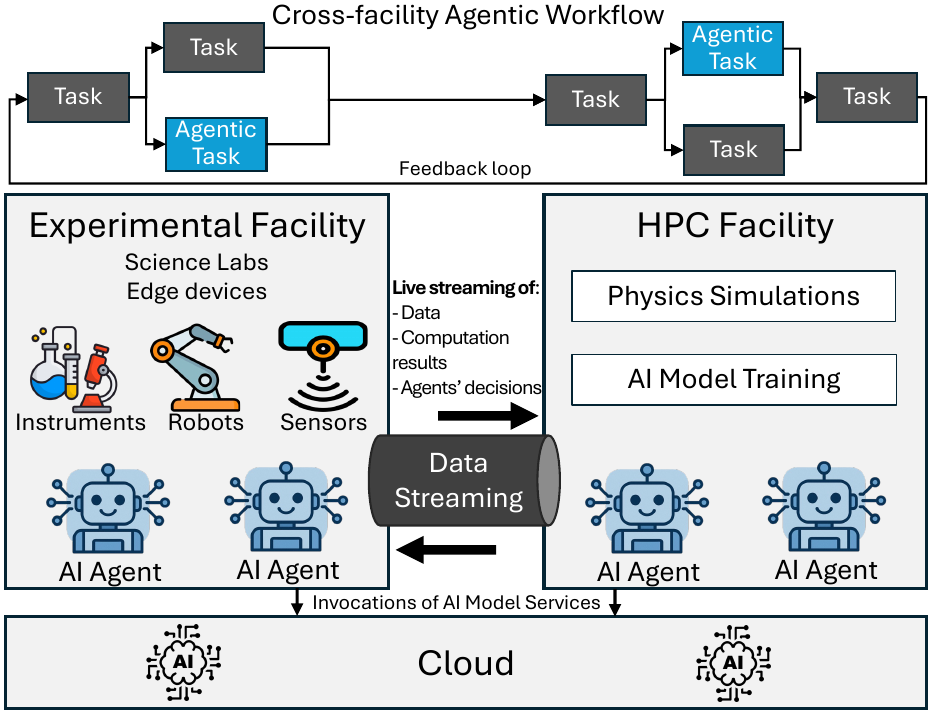}
  \caption{Agentic workflow spanning an edge-cloud-HPC continuum. Data stream in near real time from the experimental facility to HPC systems, with results feeding back into upstream tasks. Agentic tasks (tools) run alongside traditional ones, making provenance critical to trace potential hallucinations or errors that may propagate through the entire workflow.}
  \label{fig:agentic_workflows}
\end{figure}

A growing challenge in these workflows involves managing execution across physically and logically distributed facilities that include edge devices, cloud services, and HPC systems~\cite{murugesan2025rise, acharya2025agentic, souza2023towards, wf_community_summit} (Fig.~\ref{fig:agentic_workflows}). Scientific experiments may be conducted in external laboratories or at the edge, generating data in near real-time. These data must be immediately transmitted to an HPC system, where they feed into simulations, analytics pipelines, or machine learning (ML) training processes. This tight integration requires not only reliable data movement across sites but also a coherent understanding of how AI agents interact with this data across systems. 

While some MCP-based agent frameworks record prompts, responses, and AI service invocations, these data are typically isolated from the rest of the workflow. This disconnection hinders the contextualization of agent interactions or understanding their downstream impact. Existing provenance techniques lack explicit representations of key agent artifacts and their integration with the workflow. They typically model workflows as static graphs, missing the semantics needed to capture agentic behavior, dynamic decisions, and model-driven reasoning.
We argue that \textit{agentic provenance}, i.e., provenance data that track tasks executed by AI agents and their influence on downstream non-agentic tasks and data in the workflows, provides the glue power needed to unify these elements into a single, queryable graph. 
This enables traceability, root cause analysis, and continuous agent improvement, such as refining prompts or tuning model parameters to reduce hallucinations, which are essential in agentic workflows to support responsible, reproducible, and trustworthy AI-driven decisions~\cite{10678731}.

\subsection{W3C PROV and Extensions for Workflows, AI, and Agents}

\begin{figure}[htbp]
  \centering
\includegraphics[width=0.9\columnwidth]{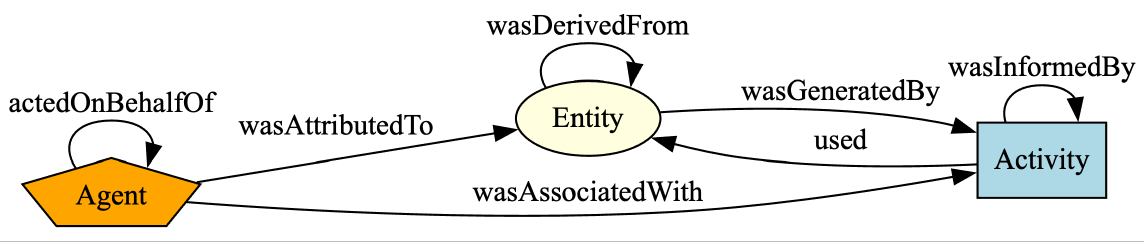}
  \caption{The W3C PROV Provenance Model~\cite{W3CPROV}.}
  \label{fig:w3c_prov}
\end{figure}

The W3C PROV standard~\cite{W3CPROV} is a widely adopted representation model for provenance, providing a structured way to describe how data were produced, by whom, and through which processes.
Fortunately, the W3C PROV standard already defines \textbf{Agent}, the central abstraction in this work, as one of its three core classes, alongside Entity (data) and Activity (process), with agents representing either software or human actors responsible for activities.
Fig.~\ref{fig:w3c_prov} shows these core classes and their main relationships among them.

PROV supports domain- and application-specific extensions and underpins many workflow provenance systems requiring standardized, interoperable representations of complex  processes. For instance, PROV-DfA~\cite{Souza2018Provenance} extends PROV to capture human actions in human-steered workflows, while ProvONE~\cite{cao2016provone} adds workflow-specific metadata and aims at supporting existing workflow management systems. For AI/ML workflows, PROV-ML~\cite{souza2022workflow} combines general workflow concepts with ML-specific artifacts, especially for model training and evaluation. FAIR4ML~\cite{fair4ml} adopts a model-centric approach to support the Findability, Accessibility, Interoperability, and Reproducibility (FAIR) principles. 
These works are orthogonal to ours, as they define complementary concepts rather than representing AI agents that steer workflows. Although the W3C PROV has been extended for agents and multi-agent systems~\cite{davis2017data, friedman2020provenance}, these earlier efforts predate agentic workflows, lacking support for core agentic AI concepts~\cite{sapkota2025ai} and how they relate to broader workflow.

\section{A Provenance Model for Agentic Workflows}
\label{sec:model}

\begin{figure}[htbp]
  \centering
  \includegraphics[width=\columnwidth]{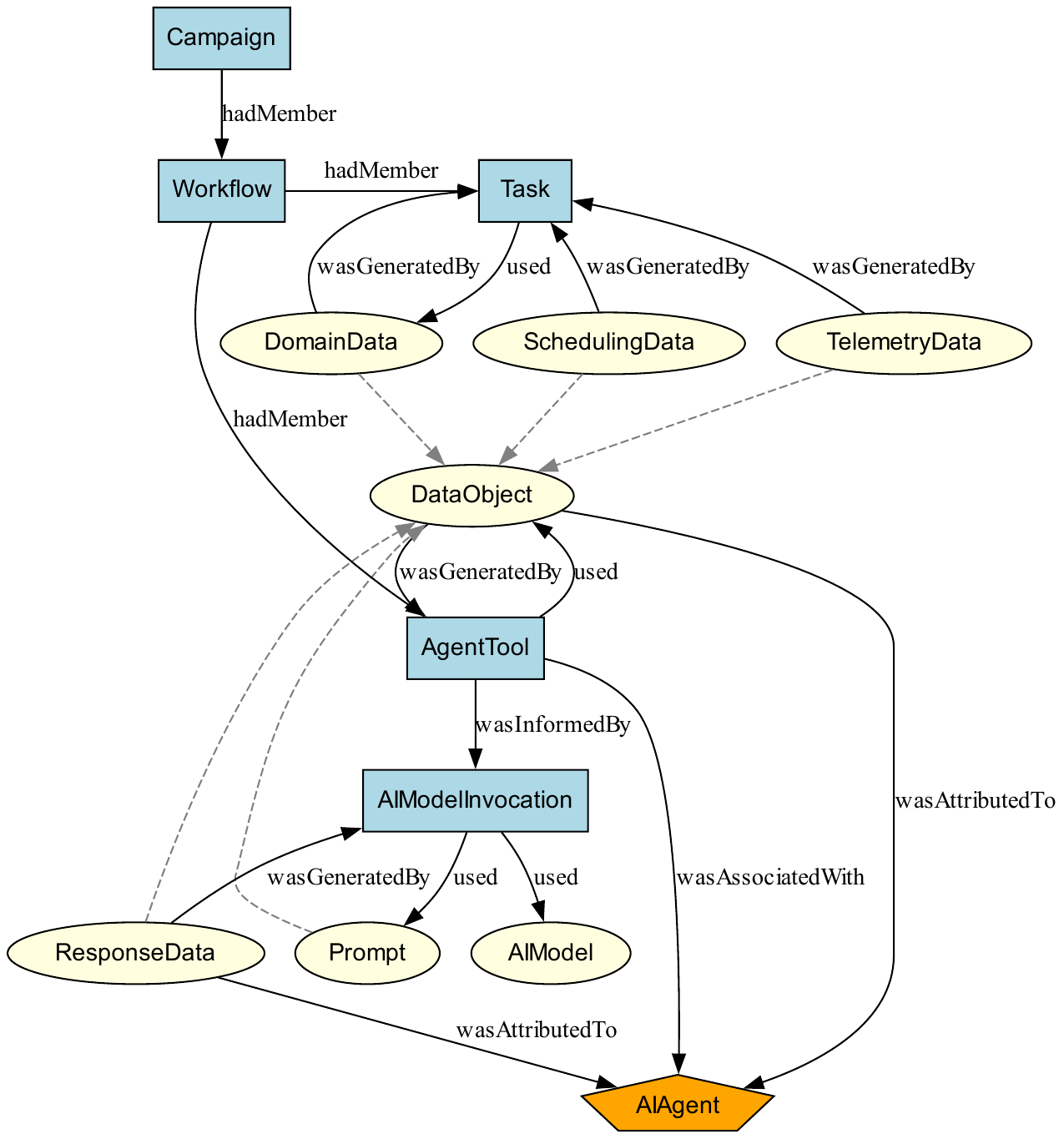}
  \caption{PROV-AGENT: A W3C PROV Extension for Agentic Workflows. Dashed arrows represent \textit{subClassOf}.}
  \label{fig:prov-agent}
\end{figure}

PROV-AGENT is a provenance model for representing AI agent interactions, model invocations, and their relationships to non-agentic tasks and data in agentic workflows (Fig.~\ref{fig:prov-agent}). It extends W3C PROV and incorporates MCP concepts to unify agents and traditional components as first-class elements.

At its core, the model includes standard workflow structures such as \class{Campaign}, \class{Workflow}, and \textit{Task}, modeled as subclasses of PROV Activities.
Campaigns are associated with Person or Organization agents via \textit{wasAssociatedWith}, omitted from the figure to reduce clutter.
Tasks consume (PROV \textit{used}) and produce (PROV \textit{generated}) domain-specific data objects (\class{DomainData}). Typically, in a provenance graph, they contain parameters, arguments, KPIs, QoIs, and pointers to domain data files or data objects stored elsewhere~\cite{provlake_escience_2019}.
Tasks also generate two additional types of metadata: \class{SchedulingData} and \class{TelemetryData}. \class{SchedulingData} contains where the task ran, including details such as compute node, CPU core, or GPU ID. \class{TelemetryData} contains runtime metrics such as CPU and GPU usage, and disk usage. These data, modeled as subclasses of \class{DataObject}, which is a subclass of PROV Entity, enrich the provenance graph with infrastructure-level context needed for traceability and performance analysis. 

We extend the abstract W3C PROV \textbf{Agent} by modeling \class{AIAgent} as its subclass, enabling a natural integration of agent actions and interactions into the broader workflow provenance graph. This modeling  is not constrained to single-agent scenarios. Multi-agentic workflows, with other AI agents, each with their own tools and reasoning paths, can be instantiated within the same provenance graph, enabling the representation of collaborative or parallel agent behaviors within a workflow.

Following the MCP terminology, an AI agent can be associated with one or many tool executions (\class{AgentTool}) and each tool may be informed by (PROV \textit{wasInformedBy}) one or many \class{AIModelInvocations}. Each \class{AIModelInvocation} uses a \class{Prompt} and a specific \class{AIModel}, which holds model metadata, including its name, type, provider, temperature, and other parameters, and \textit{generates} a \class{ResponseData} object, which is \textit{attributedTo} the corresponding agent.
Although LLMs are more common, the PROV-AGENT is designed to be modality-agnostic and supports other foundation models, such as those for vision, audio, or multimodal reasoning, as long as they follow a prompt-invocation-response interaction model.

The data used or generated by agents, including prompts, responses, are represented as subclasses of the \class{DataObject} Entity. This allows agents to consume and produce not only \class{DomainData}, but also system-level and contextual data such as \class{SchedulingData} and \class{TelemetryData}.
When instances of subclasses of \class{DataObject} are generated by agent tools, they are attributed to (\textit{wasAttributedTo}) the instance of the agent.
The additional data types can be used by the agent as part of reasoning or planning, for example through RAG strategies to enhance prompts with relevant contextual knowledge. Since the relationships are explicitly modeled using standard PROV constructs such as \textit{used}, \textit{wasGeneratedBy}, \textit{wasAssociatedWith}, and \textit{wasInformedBy}, the resulting graph is fully connected and queryable. This enables users to trace a final output or decision all the way back through agent reasoning, prompts, input data, system context, and execution metadata, addressing the key challenge of capturing agentic behavior as part of end-to-end workflows.

\section{System Implementation and Evaluation}
\label{sec:validation}

\subsection{Implementation}

Rather than building a new provenance system from scratch, we extend \textit{Flowcept}~\cite{flowcept}, an open-source distributed provenance framework designed for complex, heterogeneous workflows spanning experimental facilities at the edge, cloud platforms, and HPC environments. Flowcept uses a federated, broker-based model where raw provenance data, which may come in varied formats and schemas, can be streamed from instrumented scripts, data observability hooks in workflow tools (e.g., Dask, MLflow), and from data streaming services and storage layers such as Redis, Kafka, SQLite, file systems, and object stores while the workflows run~\cite{souza2023towards}.
A central consolidation service unifies, curates this data into a persistent provenance database while applying a W3C PROV-extended model, making it well-suited for capturing and contextualizing provenance in end-to-end agentic workflows.

Building on the MCP concepts, when the MCP server is initialized, we begin by creating a new instance of \class{AIAgent}, assigning it an identifier and name so it can be properly associated with the tools it executes. Flowcept supports several instrumentation methods, but for MCP-based tools, the  most straightforward method is via decorators. In generic Python functions, applying the \texttt{\char64 flowcept\_task} decorator ensures that, upon execution, the function's inputs, outputs, and any generated telemetry or scheduling data are automatically captured. Follow this approach, since MCP tools have well-defined input arguments and return values, we introduce a new decorator, \texttt{\char64 flowcept\_agent\_tool}, which creates a corresponding \class{AgentTool} execution activity for each tool execution. This activity is associated with the executing agent and linked to its inputs and outputs using the PROV relationships defined in the PROV-AGENT model.

Tool executions are often informed by one or more \class{AIModel} calls. Given our driving use cases and that most agentic workflow users employ LLMs for their agents, this first implementation of PROV-AGENT focuses on supporting LLMs by providing a generic wrapper for abstract LLM objects, compatible with models from popular LLM interfaces, including CrewAI, LangChain, and OpenAI. Whenever a prompt is sent to an LLM service provider in the cloud (e.g., OpenAI, SambaNova, Azure), the wrapper captures the prompt, response, model metadata (e.g., provider, name, and parameters like temperature), and optional telemetry such as response time. Fig.~\ref{lst:agent_tool} shows an MCP tool example annotated with the decorator and using the wrapper \texttt{FlowceptLLM}.  
Model metadata are recorded within an instance of \class{AIModel} and each invocation is recorded as an \class{AIModelInvocation} activity and linked to the model, prompt, and response, according to the defined relationships. When a tool depends on LLM results, Flowcept establishes a \textit{wasInformedBy} relationship from the \class{AgentTool} to the relevant \class{AIModelInvocation} activities. While this implementation records only the agent's ID and name, the model supports extended metadata, such as model and tools' version control state, and further configuration parameters.



\begin{figure}[ht]
\begin{lstlisting}
from langchain_openai import ChatOpenAI
from flowcept import FlowceptLLM, flowcept_agent_tool

@mcp.tool()
@flowcept_agent_tool
def evaluate_scores(layer, result, scores): 
    ...
    prompt = get_prompt(layer, result, scores)
    llm = FlowceptLLM(ChatOpenAI(model="gpt-4o"))
    response = llm.invoke(prompt)
    ...
    return ...
\end{lstlisting}
\caption{MCP agent tool that invokes an LLM to assess physics model outputs. With the decorator \texttt{@agent\_flowcept\_task} and \texttt{FlowceptLLM} wrapper, agent tool and LLM invocation provenance are captured.}
\label{lst:agent_tool}
\end{figure}

Flowcept also provides an MCP agent with a Streamlit GUI that enables users to interact with the provenance database through natural language queries at runtime. While the details of this agent are beyond the scope of this paper, in the next section we highlight how it helps users to query and explore the provenance data captured using PROV-AGENT.

\subsection{Preliminary Evaluation}

In this section, we evaluate PROV-AGENT and its implementation by demonstrating how agent decisions, LLM interactions, and workflow tasks are unified in a single provenance graph, enabling users to trace erroneous outputs back to their upstream prompts, inputs, and prior decisions.

\textbf{Use case.}
We employ PROV-AGENT in an autonomous additive manufacturing workflow being developed at Oak Ridge National Laboratory (ORNL)~\cite{aam_website}.
This envisioned workflow integrates a metal 3D printer at ORNL's Manufacturing Demonstration Facility (MDF) on the Edge with an HPC system at the ORNL Leadership Computing Facility (OLCF), streaming sensor data in near real-time to HPC simulations, illustrating a concrete case of the generic workflow in Fig.~\ref{fig:agentic_workflows}. Although the direct live data connection between the sensors and simulation is still under development, our implementation already applies to the agentic control loop and distributed facilities at ORNL.
At MDF, sensor drivers collect data layer by layer as a metal component is fabricated. This layer-specific data are used to estimate the current state of the system.
Using the approach of model predictive control~\cite{model_ref_ctrl_review}, a forward-looking physics-based model explores the downstream consequences of decisions for upcoming layers. Each prospective decision for the upcoming layer is scored using an analysis routine. 
Researchers are investigating the benefits of using AI-driven decision-making via Analysis Agent tools invoking an LLM (\textit{gpt-4o}) service hosted in the cloud.
The agents use structured prompts to decide which control result is best for print control based on their scores and other data in the agent context, such as previous decisions and user guidance. 
Thus, the decision made for each layer informs the decision logic in the next, enabling the system to learn over the course of a print.
However, because the agent relies on an LLM, there is a risk of hallucinated or incorrect outputs. Since each decision influences the next in this iterative loop, a single error may propagate across layers, potentially compromising downstream outputs, thus making provenance tracking essential.

\begin{figure*}
  \centering
  \includegraphics[width=1.0\textwidth]{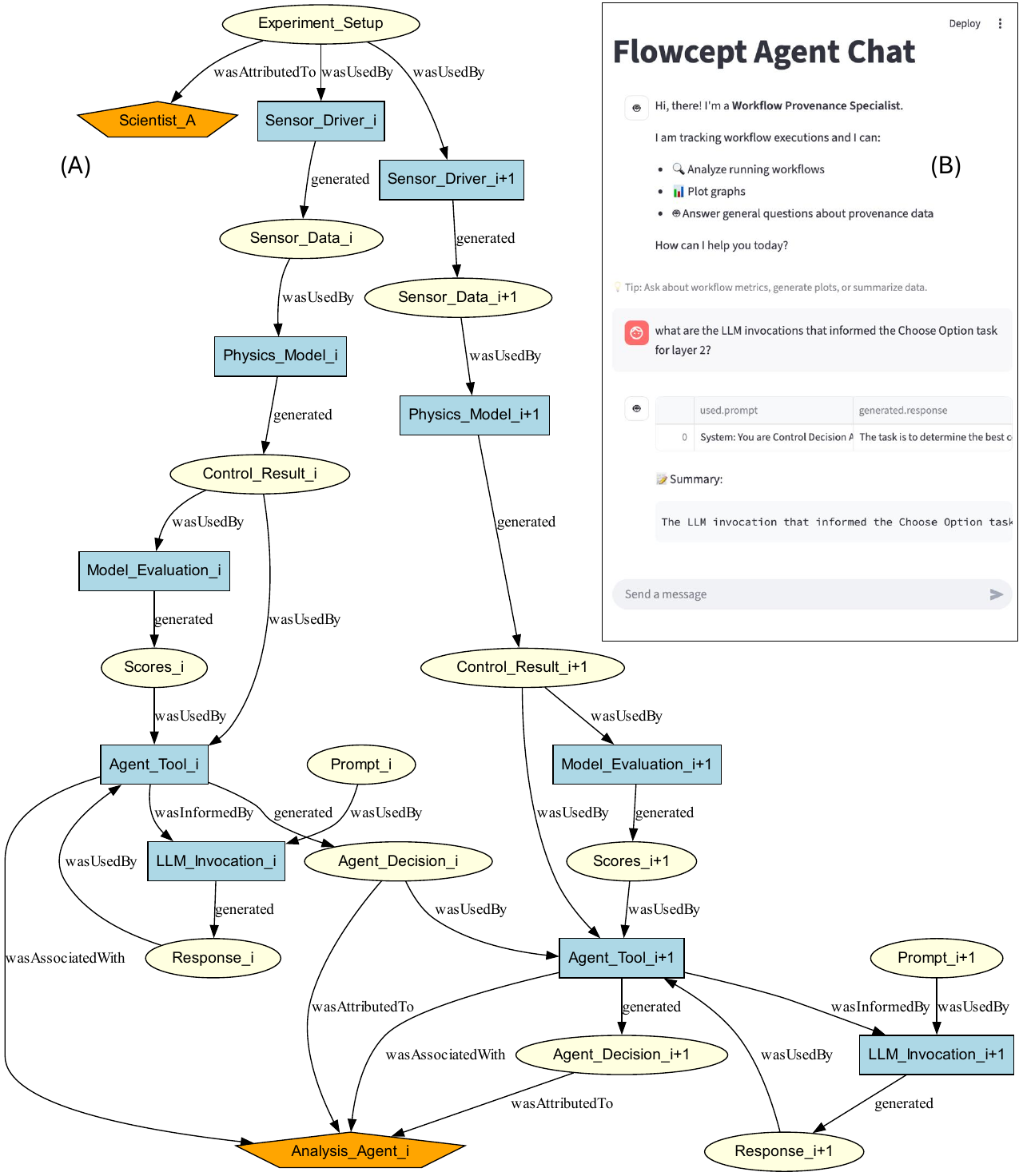}
 \caption{(A) Instantiation of the unified provenance graph using PROV-AGENT for an additive manufacturing workflow. Sensor drivers run on edge devices, while agents and physics models run on HPC. Sensor data are generated layer by layer and used by the simulation model to assess print quality. An AI agent analyzes results and makes layer-specific decisions, where decisions at iteration $i$ influence iteration $i{+}1$. Only two iterations are shown, though typical workflows span up to thousands. Arrows reverse standard W3C PROV directions (\textit{used}, \textit{wasGeneratedBy}) for top-down clarity.\\
(B) Chat window showing a natural language query to the Flowcept Agent for Query 3.}
  \label{fig:prov-graph}
\end{figure*}

\textbf{End-to-end Provenance Graph.} 
Figure~\ref{fig:prov-graph}-A shows how PROV-AGENT would function in the additive manufacturing use case. After the scientist inputs the experiment setup, the driver (\inst{Sensor\_Driver\_i}) iteratively triggers the sensors for each printed layer $i$. The resulting \inst{Sensor\_Data\_i} is streamed to the HPC system for processing by a physics-based model (\inst{Physics\_Model\_i}) and evaluation task (\inst{Model\_Evaluation\_i}), producing control results and scores. \inst{Experiment\_Setup}, \inst{Sensor\_Data\_i}, \inst{Control\_Result\_i}, and \inst{Scores\_i} are modeled as \class{DomainData}, and their linked activities are \class{Tasks}.

For every layer $i$, an agent decision-making tool is executed to assess the scores for physics model predictions, creating an instance (\inst{Agent\_Tool\_i}) of the class \class{AgentTool},  and is linked to an instance of the \class{AIAgent}, the \inst{Analysis\_Agent\_i}. Every tool execution for layer $i$ uses the outputs of the physics model and their evaluation scores, and invokes the cloud-based LLM models via \inst{LLM\_Invocation\_i}, which are instances of \class{AIModelInvocation}. LLM invocations are explicitly connected to the used \inst{Prompt\_i} and generated \inst{Response\_i} instances, where the responses are attributed to the agent instance. The resulting agent decisions (\inst{Agent\_Decision\_i}), which are instances of \class{DomainData}, are also attributed to the agent, completing each agentic reasoning cycle. 

Several model components from the PROV-AGENT schema are intentionally omitted from the figure, including \class{Campaign}, \class{Workflow}, \class{TelemetryData}, and \class{SchedulingData}. These classes are recorded in the underlying provenance database but excluded from the visual representation to reduce clutter. Activities, e.g., workflow tasks and agent tools, are linked to location metadata indicating where they ran (e.g., edge, HPC, or cloud). Though omitted in the figure, these PROV Location entities help map execution across the Edge–Cloud–HPC continuum.

\textbf{Query examples enabled by PROV-AGENT.} With PROV-AGENT, several new queries are enabled to support agent accountability and tracing back when errors/hallucinations happen. Below we show a few examples of queries using a distributed Edge-Cloud-HPC workflow that mimics the agentic additive manufacturing workflow under development.

\begin{itemize}[left=0pt, nosep]

  \item \textbf{Q1. Given an agent decision, what was the complete lineage until the first input data?} \\
  Given an agent decision \inst{Agent\_Decision\_i}, the query traverses to its generating \inst{Agent\_Tool\_i}, then to the inputs it used: \inst{Scores\_i}, \inst{Control\_Result\_i}, and \inst{Agent\_Decision\_{i-1}}. These are traced back through \inst{Model\_Evaluation\_i} and \inst{Physics\_Model\_i} to the original \inst{Sensor\_Data\_i} that was generated by the driver for layer $i$ when utilized the recorded \inst{Experiment\_Setup}.

  \item \textbf{Q2. When printing layer 2, what was the agent decision, the available score options, and the reasoning behind the decision?} \\
  Starting at \inst{Agent\_Decision\_2}, we trace back to \inst{Agent\_Tool\_2} and inspect the input \inst{Scores\_2}, \inst{Control\_Result\_2}, and the \inst{Response\_2} from the  \inst{LLM\_Invocation\_2} to understand the reasoning context.

  \item \textbf{Q3. What was the LLM prompt and response when a surprising agent decision was identified?} \\
  Given that a hallucination was identified when the agent was deciding on the scores for layer 2, after identifying the unexpected \inst{Agent\_Decision\_2}, the query traces back to \inst{Agent\_Tool\_2} and its linked \inst{LLM\_Invocation\_2} to retrieve the corresponding \inst{Prompt\_2} and \inst{Response\_2}. This query is illustrated in the Streamlit chat GUI of Flowcept agent in Figure~\ref{fig:prov-graph}-B.

  \item \textbf{Q4. How did an agent decision influence subsequent workflow activities?} \\
  Given that an agent decision \inst{Agent\_Decision\_i} is used by another \inst{Agent\_Decision\_{i+1}}, the query recursively navigates on the used/wasGeneratedBy relationships in the path between the \inst{Agent\_Decision\_i} and the \inst{Agent\_Decision} in the last layer.

  \item \textbf{Q5. Where did erroneous data originate, and how did it propagate?} \\
    After identifying a faulty \inst{Agent\_Decision\_i}, the query traces backward through the tool, LLM response, model outputs, and \inst{Sensor\_Data\_i} to find the cause, and forward to identify affected downstream results.

\end{itemize}

These queries demonstrate how PROV-AGENT enables end-to-end analysis of agent behavior within workflows, supporting accountability, debugging, and iterative improvement.

\section{Conclusion}

As AI agents become core components of workflows, ensuring transparency, accountability, and reproducibility is critical, especially given their non-deterministic behavior and potential to hallucinate and propagate errors across data and tasks in the workflows. 
PROV-AGENT addresses this need by extending the W3C PROV standard and leveraging the Model Context Protocol (MCP) to capture fine-grained agentic provenance. We build on established provenance foundations to add key AI agent artifacts, including tools, prompts, responses, and model invocations, and integrate them to the non-agentic tasks and data in the workflow.
This unified approach not only supports traceability and root cause analysis but also enables continuous agent improvement through the comparison of decisions across prompt engineering refinements or fine-tuning of model parameters.
We demonstrated our open-source implementation in an agentic workflow use case running on distributed facilities with feedback loops and AI agents steering execution.
To the best of our knowledge, this is the first provenance framework designed to track agent actions and decisions in agentic workflows. While this is an early step, it establishes a foundation that researchers and practitioners can build on, not only to enable root-cause analysis, interpretability, and model and prompt fine-tuning, but also to explore new techniques for detecting and ultimately remediating hallucinations in AI-driven decisions.

\medskip
{\small
\noindent \textbf{Acknowledgments.}
The authors thank the ORNL team: 
Miaosen Chai,
Timothy Poteet,
Phillipe Austria,
Marshall McDonnell,
Ross Miller,
A.J. Ruckman,
Tyler Skluzacek,
Feiyi Wang, 
Sarp Oral,
Arjun Shankar
for their help with the use case development.
ChatGPT-4o was used to help polish writing, improve conciseness, and check grammar. 
This research used resources of the Oak Ridge Leadership Computing Facility at ORNL, supported by the Office of Science, U.S. Department of Energy (DOE) under Contract No.  DE-AC05-00OR22725. Additional DOE support was provided under Contract No. DE-AC02-06CH11357.
}


\bibliographystyle{IEEEtran}
\bibliography{references}

\end{document}